# Statistical modelling of primary Ewing tumours of the bone

**Sreepurna Malakar**
**Department of Chemistry and Biochemistry**
**University of Alaska, U.S.A.**

**Florentin Smarandache**
**Department of Mathematics and Statistics**
**University of New Mexico, U.S.A.**

**Sukanto Bhattacharya**
**Department of Business Administration**
**Alaska Pacific University, U.S.A.**

**Abstract**

This short technical paper advocates a bootstrapping algorithm from which we can form a statistically reliable opinion based on limited clinically observed data, regarding whether an osteo-hyperplasia could actually be a case of Ewing's osteosarcoma. The basic premise underlying our methodology is that a primary bone tumour, if it is indeed Ewing's osteosarcoma, cannot increase in volume beyond some critical limit without showing metastasis. We propose a statistical method to extrapolate such critical limit to primary tumour volume. Our model does not involve any physiological variables but rather is entirely based on time series observations of increase in primary tumour volume from the point of initial detection to the actual detection of metastases.

## I. Introduction

To date, oncogenetic studies of *EWS/FLI-1*1 induced malignant transformation have largely relied upon experimental manipulation of Ewing's bone tumour cell lines and fibroblasts that have been induced to express the oncogene. It has been shown that the biology of Ewing's tumour cells *in vitro* is dramatically different between cells grown as mono-layers and cells grown as anchorage-independent, multi-cellular spheroids (MCS). The latter is more representative of primary Ewing's tumour *in vivo* (Lawlor et. al, 2002).

MCS are clusters of cancer cells, used in the laboratory to study the early stages of avascular tumour growth. Mature MCS possess a well-defined structure, comprising a central core of necrotic i.e. dead cells, surrounded by a layer of non-proliferating, quiescent cells, with proliferating cells restricted to the outer, nutrient-rich layer of the tumour. As such, they are often used to assess the efficacy of new anti-cancer drugs and treatment therapies. The majority of mathematical models focus on the growth of MCS or avascular tumour growth. Most recent works have focused on the evolution of MCS growing in response to a single, externally-supplied nutrient, such as oxygen or glucose, and usually two growth inhibitors.

Mathematical models of MCS growth typically consist of an *ordinary differential equation* (ODE) coupled to one or more *reaction-diffusion equations* (RDEs). The ODE is derived from mass conservation and describes the evolution of the outer tumour boundary, whereas the RDEs describe the distribution within the tumour of vital nutrients such as oxygen and glucose and growth inhibitors (Dorman and Deutsch, 2002). However studies of this type, no matter how mathematically refined, often fall short of direct clinical applicability because of rather rigorous restrictions imposed on the boundary conditions. Moreover, these models focus more on the structural evolution of a tumour that is already positively classified as cancerous rather than on the clinically pertinent question of whether an initially benign growth can at a subsequent stage become invasive and show metastases (De Vita et. al., 2001).

What we therefore aim to devise in our present paper is a bootstrapping algorithm from which we can form an educated opinion based on clinically observed data, regarding whether a bone growth initially diagnosed as benign can subsequently prove to be malignant (i.e. specifically, a case of Ewing's osteosarcoma) . The strength of our proposed algorithm lies mainly in its computational simplicity – our model does not involve any physiological variables but is entirely based on time series observations of progression in tumour volume from the first observation point till detection of metastases.

## II. Literature support

In a clinical study conducted by Hense et. al. (1999), restricted to patients with suspected Ewing's sarcoma, tumour volumes of more than 100 ml and the presence of primary metastases were identified as determinants of poor prognosis in patients with such tumours. Diagnoses of primary tumours were ascertained exclusively by biopsies. The diagnosis of primary metastases was based on *thoracic computed tomography* or on whole body bone scans. It was observed that of 559 of the patients (approx. 68% in a total sample size of 821) had a volume above 100 ml with smaller tumours being more common in childhood than in late adolescence and early adulthood. Extensive volumes were observed in almost 90% of the tumours located in femur and pelvis while they were less common in other sites ($p < 0.001$). On average, 26% of all patients were detected with clinically apparent primary metastases.

The detection rate of metastases was markedly higher in patients diagnosed after 1991 ($p < 0.001$). Primary metastases were also significantly more common for tumours originating in the pelvis and for other tumours in the Ewing's family of tumours (EFT); mainly the peripheral neuro-ectodermal tumours (PNET); ($p < 0.01$). Tumours greater than 100 ml were positively associated with metastatic disease ($p < 0.001$). Multivariate analyses, which included simultaneously all univariate predictors in a *logistic regression model*, indicated the observed associations were mostly unconfounded.

Further it has been found that the metastatic potential of human tumours is encoded in the bulk of a primary tumour, thus challenging the notion that metastases arise from sparse cells within a primary tumour that have the ability to metastasize (Sridhar Ramaswamy et. al., 2003). These studies lend credence to our fundamental premise about a critical primary tumour volume being used as a classification factor to distinguish between benign and potentially malignant bone growth.

## III. Statistical modelling methodology

Assuming that the temporal drift process governing the progression in size of a primary Ewing tumour of the bone to be linear (the computationally simplest process), we suggest a straightforward computational technique to generate a large family of possible tumour propagation paths based on clinically observed growth patterns under laboratory conditions. In case the governing process is decidedly non-linear, then our proposed scheme would not be applicable and in such a case one will have to rely on a completely non-parametric classification technique like e.g. an Artificial Neural Network (ANN).

Our proposed approach is a bootstrapping one, whereby a linear autoregression model is fitted through the origin to the observation data in the first stage. If one or more beta coefficients are found to be significant at least at a 95% level for the fitted model then, in the second stage, the autoregression equation is formulated and solved as a *linear difference equation* to extract the governing equation.

In the final stage, the governing equation obtained as above is plotted, for different values of the constant coefficients, as a family of possible temporal progression curves generated to explain the propagation property of that particular strain of tumour. The critical volume of the primary growth can thereafter be visually extrapolated from the observed cluster of points where the generated family of primary tumour progression curves show a *definite uptrend* vis-a-vis the actual progression curve.

If no beta coefficient is found to be significant in the first stage, a non-linear temporal progression process is strongly suspected and the algorithm terminates without proceeding onto the subsequent stages, thereby implicitly recommending the problem to a non-parametric classification model.

The mathematical structure of our proposed model may be given as follows:

Progression in primary Ewing tumour size over time expressed as an n-step general autoregressive process through the origin:

$$S_t = \sum_{j=1}^{n} \beta_j S_{t-j} + \varepsilon \qquad \text{(I)}$$

Formulated as a linear, difference equation we can write:

$$-S_t + \beta_1 S_{t-1} + \beta_2 S_{t-2} + \ldots + \beta_n S_{t-n} = -\varepsilon \qquad \text{(II)}$$

Taking $S_t$ common and applying the *negative shift operator* throughout, we get:

$$[-1 + \beta_1 E^{-1} + \beta_2 E^{-2} + \ldots + \beta_n E^{-n}] S_t = -\varepsilon \qquad \text{(III)}$$

Now applying the *positive shift operator* throughout we get:

$$[-E^n + \beta_1 E^{n-1} + \beta_2 E^{n-2} + ... + \beta_n] S_t = -\varepsilon \qquad \text{(IV)}$$

The *characteristic equation* of the above form is then obtained as follows:

$$-r^n + \beta_1 r^{n-1} + \beta_2 r^{n-2} + ... + (\beta_n + \varepsilon) = 0 \qquad \text{(V)}$$

Here r is the root of the characteristic equation. After solving for r, the governing equation can be derived in accordance with the well-known analytical solution techniques for ordinary linear difference equations (Kelly and Peterson, 2000).

**IV. Simulated clinical study**

We set up a simulated clinical study applying our modelling methodology with the following hypothetical primary Ewing tumour progression data adapted from the clinical study of Hense et. al. (1999) as given in Table I below:

**Table I**

| Observation (t) | Primary Ewing tumour volume (in ml.) |
|---|---|
| (At point of first detection)<br><br>1 | <br><br>5 |
| 2 | 7 |
| 3 | 9 |
| 4 | 19 |
| 5 | 39 |
| 6 | 91 |
| 7<br><br>(At the point of detection of metastasis) | 102 |

**Figure I**

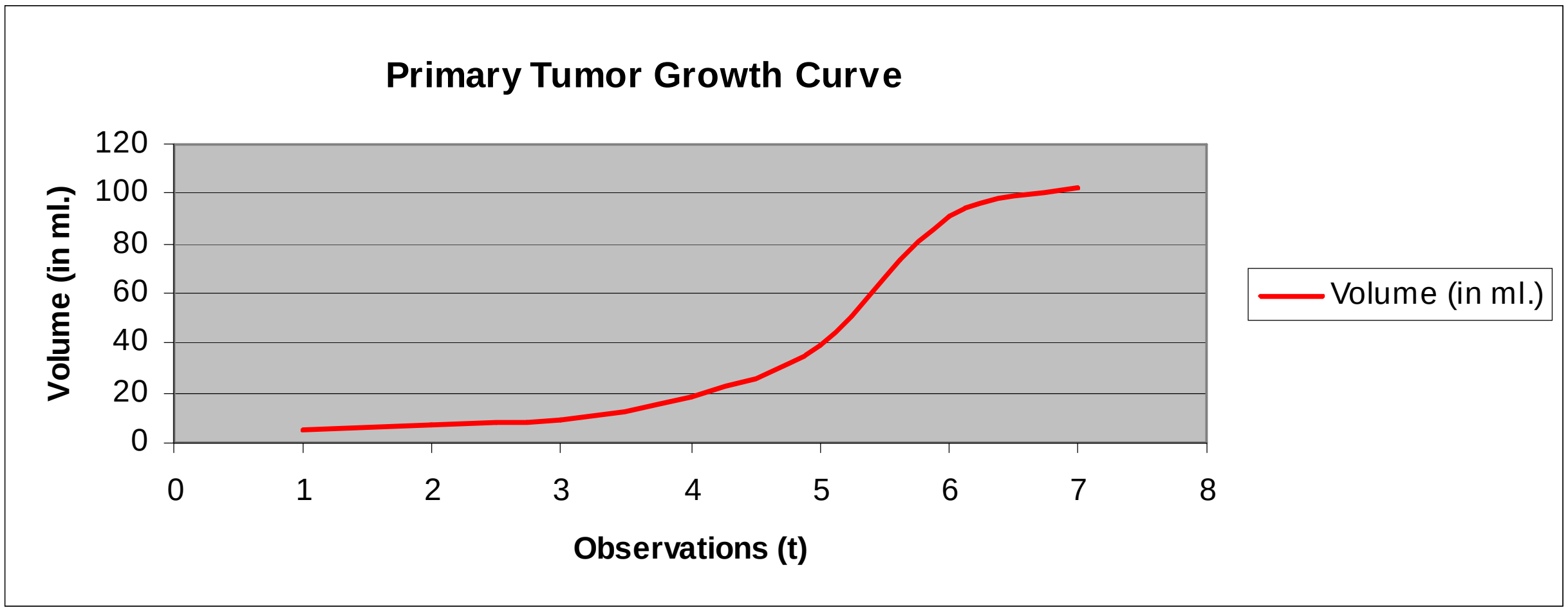

The temporal progression path of the primary growth from the point of first detection to the onset of metastasis is plotted above in Figure I.

We have fitted an AR (2) model to the primary tumour growth data as follows:

$$E\,(S_t) = -1.01081081S_{t-1} + 5.32365561S_{t-2} \qquad \text{(VI)}$$

The $R^2$ of the fitted model is approximately 0.8311 and the *F*-statistic is 9.83832 with an associated p-value of approximately 0.04812. Therefore the fitted model definitely has an overall predictive utility at the 5% level of significance.

The residuals of the above AR (2) fitted model are given in Table II as follows:

**Table II**

| ***Observation*** | ***Predicted $S_t$*** | ***Residuals*** |
|---:|---:|---:|
| 1 | -5.05405405 | 12.05405405 |
| 2 | 19.5426024 | -10.5426024 |
| 3 | 28.168292 | -9.168292003 |
| 4 | 28.7074951 | 10.29250488 |
| 5 | 61.7278351 | 29.27216495 |
| 6 | 115.638785 | -13.63878518 |

The average of the residuals comes to 3.044841. Therefore the linear difference equation to be solved in this case is as follows:

$$X_t = -1.01081081X_{t-1} + 5.32365561X_{t-2} + 3.044841 \qquad \text{(VII)}$$

Applying usual solution techniques, the general solution to equation (VII) is obtained as follows:

$$X_t = c_1 (2.43124756)^t + c_2 (-3.44205837)^t \quad \text{(VIII)}$$

Here $c_1$ and $c_2$ are the constant coefficients which may now be suitably varied to generate a family of possible primary tumour progression curves as in Figure II below:

**Figure II**

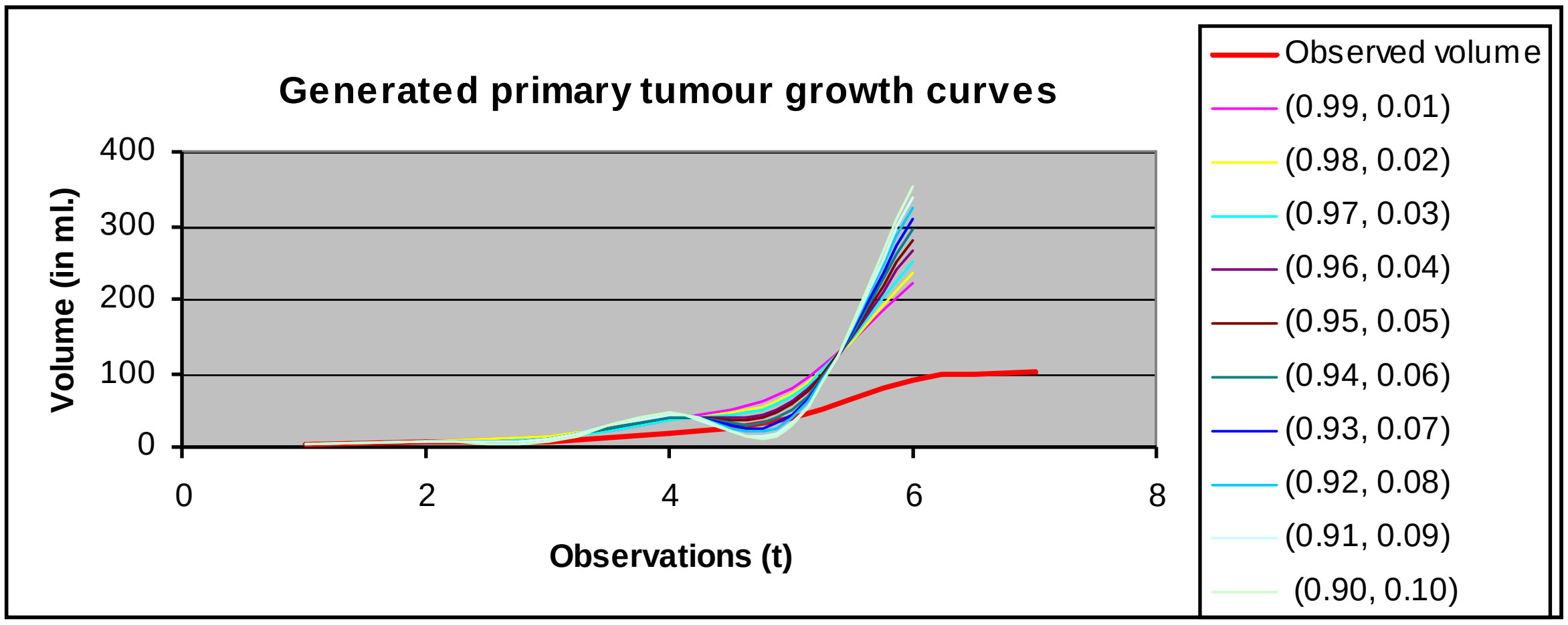

In the above plot, we have varied $c_2$ in the range 0.01 to 0.10 and imposed the condition $c_1 = 1 - c_2$. The other obvious condition is that choice of $c_1$ and $c_2$ would be such as to rule out any absurd case of negative volume. Of course the choice of the governing equation parameters would also depend on specific clinical considerations (King, 2000).

## V. Conclusion

From Figure II, it becomes visually apparent that continuing increase in the observed size of the primary growth beyond approximately 52 ml. in volume would be potentially malignant as this would imply that the tumour would possibly keep exhibiting uncontrolled progression till it shows metastasis. This could also be obtained arithmetically as the average volume for $t = 5$. Therefore the critical volume could be fixed around 52 ml. as per the computational results obtained in our illustrative example.

Though our computational study is intended to be purely illustrative as we have worked with hypothetical figures and hence cannot yield any clinical conclusion, we believe we have hereby aptly demonstrated the essential algorithm of our statistical approach and justified its practical usability under laboratory settings. We have used a difference equation model rather than a differential equation one because under practical laboratory settings, observations cannot be made continuously but only at discrete time intervals. There is immediate scope of taking our line of research further forward by actually implementing an autoregressive process to model *in vitro* growth of MCS with real data.

***Journals and text references:***

***Website references:***